\documentclass[namedreferences,hyperref,optionalrh]{spr-sola}
\usepackage{graphicx}        
\usepackage{color}           

\chardef\us=`\_

\begin{document}

\begin{frontmatter}
\title{Bridging Solar and Stellar Physics: Role of SDO in Understanding Stellar Active Regions and Atmospheric Heating}

\author[addressref={aff1},email={toriumi.shin@jaxa.jp}]{\inits{S.}\fnm{Shin}~\snm{Toriumi}\orcid{0000-0002-1276-2403}}
\address[id=aff1]{Institute of Space and Astronautical Science, Japan Aerospace Exploration Agency, 3-1-1 Yoshinodai, Chuo-ku, Sagamihara, Kanagawa 252-5210, Japan}

\runningauthor{Toriumi}
\runningtitle{Bridging Solar and Stellar Physics}

\begin{abstract}
The solar-stellar connection provides a unique framework for understanding magnetic activity and atmospheric heating across a broad spectrum of stars. Solar Dynamics Observatory (SDO) of NASA, equipped with the Helioseismic and Magnetic Imager, Atmospheric Imaging Assembly, and Extreme ultraviolet Variability Experiment, has enabled detailed Sun-as-a-star studies that bridge solar and stellar physics. Integrating spatially resolved solar observations into disk-integrated datasets, these studies provide insights into magnetic activity occurring in distant stars. This review highlights key results from recent analyses that employed all three SDO instruments to characterize active regions, quantify universal heating relationships, and reconstruct stellar X-ray and ultraviolet spectra. We discuss how these findings advance our understanding of stellar magnetic activity, provide predictive tools for exoplanetary environments, and outline future directions for applying solar-based frameworks to diverse stellar populations.
\end{abstract}
\end{frontmatter}

\section{Introduction}\label{sec:introduction}

The solar-stellar connection is a rapidly advancing field of research within astronomy, significant for linking detailed solar observations with the diversity of stellar activity. As the closest star to us, the Sun provides a unique opportunity to study the physical processes that govern stellar magnetism, atmospheric heating, wind acceleration, and flare eruptions with unparalleled spatial and temporal resolution. These processes, i.e., the emergence of magnetic flux, energy accumulation, and release, form the foundation of our understanding of stellar magnetism and its impact on planetary environments \citep[e.g.,][]{2007LRSP....4....3G,2012LRSP....9....1R}.

Cool dwarf stars with spectral types of F, G, K, and M, including the Sun, with outer convection envelopes, host dynamo-generated magnetic fields and possess hot outer atmospheres. The mechanism that heats stellar coronae to temperatures of 10$^{6}$--10$^{7}$ K is one of the most persistent unsolved issues in astrophysics \citep[e.g.,][]{2004A&ARv..12...71G,2006SoPh..234...41K,2015RSPTA.37340259T}. Recent observations have revealed that Sun-like stars produce superflares that are more than 10 times stronger than solar flares \citep[e.g.,][]{2012Natur.485..478M,2021ApJ...906...72O,2024LRSP...21....1K,2024Sci...386.1301V}. Therefore, how active regions emerge, evolve, and undergo magnetohydrodynamic processes to produce massive explosions is a key mystery common to both the Sun and Sun-like stars \citep[e.g.,][]{2011LRSP....8....6S,2015LRSP...12....1V,2017LRSP...14....4B,2019LRSP...16....3T,2021LRSP...18....5F,2022AdSpR..70.1549T}. In this review, we use the term {\it Sun-like stars} to include F--K dwarfs with outer convection zones and dynamo-driven magnetic activity as this reflects the continuity of magnetic activity scaling relations across these spectral types.

The importance of solar-stellar studies goes beyond merely comparing stars. Stellar magnetic activity influences mass-loss rates, angular momentum loss, and even the habitability of exoplanets through space weather and space climate \citep[e.g.,][]{2014MNRAS.441.2361V,2016NatGe...9..452A,2020IJAsB..19..136A,2019LNP...955.....L}. Using the Sun as a template enables us to test theories of magnetic heating, wind acceleration, and eruption, which can be extrapolated to stars of different masses, rotation rates, and ages. This approach is valuable because the Sun is the only star that can be observed in spatial detail, enabling direct measurements of radiation output across different magnetic environments and atmospheric layers. These observational data serve as a benchmark for interpreting unresolved stellar data and validating theoretical models of stellar activity.

NASA's Solar Dynamics Observatory \citep[SDO;][]{2012SoPh..275....3P} has played a critical role in advancing solar-stellar studies. Since its launch in 2010, SDO has provided continuous, high-resolution observations of the Sun's photosphere, chromosphere, and corona through instruments called the Helioseismic and Magnetic Imager \citep[HMI;][]{2012SoPh..275..207S,2012SoPh..275..229S}, Atmospheric Imaging Assembly \citep[AIA;][]{2012SoPh..275...17L}, and Extreme ultraviolet Variability Experiment \citep[EVE;][]{2012SoPh..275..115W}. These instruments provide vector magnetic field measurements, multi-wavelength imaging, and extreme ultraviolet (EUV) spectra, enabling in-depth analysis of active regions, coronal loops, and small-scale heating events. Such data are indispensable for developing scaling laws and theoretical models applicable to other stars.

Solar observations offer unique advantages for studying magnetic activity and atmospheric heating. The magnetograms of HMI enable us to monitor the dynamic evolution of photospheric magnetic fields in flare-productive active regions from birth to eruption. The high spatial resolution images of AIA resolve coronal structures of active regions down to sub-arcsec scales, and its temporal resolution of 10s of seconds enables capturing dynamic processes such as magnetic reconnection and wave propagation. The full-disk EUV measurements of EVE offer precise monitoring of solar variability that is critical for understanding coronal heating and its impact on space weather. These capabilities enable us to quantify energy release rates, plasma flow, and magnetic field evolution in unprecedented detail. However, extrapolating solar results to stellar contexts introduces challenges. The Sun occupies a narrow region of stellar parameter space: a G2V star with moderate activity and slow rotation. However, stellar activity spans a wide range of rotation periods, magnetic field strengths, and convection zone depths, leading to diverse manifestations of magnetic heating \citep[e.g.,][]{2014ApJ...794..144R}. Furthermore, although solar observations are spatially resolved, stellar observations are disk-integrated, making isolation of contributions from individual structures challenging. This discrepancy necessitates developing empirical scaling laws and statistical models that bridge the gap between resolved solar structures and unresolved stellar signals \citep[e.g.,][]{2000ssma.book.....S}.

Recent decades have witnessed remarkable progress in stellar activity studies, driven by space-based photometry and high-energy spectroscopy. Telescopes such as Kepler \citep{2010Sci...327..977B} and Transiting Exoplanet Survey Satellite \citep[TESS;][]{2015JATIS...1a4003R} have enabled precise monitoring of stellar brightness variations, revealing rotational modulations due to starspots and impulsive brightenings due to flares \citep[e.g.,][]{2013ApJ...769...37B,2012Natur.485..478M,2013ApJ...771..127N,2016ApJ...829...23D,2023ApJ...948...64I}. These datasets provide constraints on spot coverage, lifetimes, differential rotation, and eruptivity, which are key indicators of magnetic activity. Complementary observations in X-ray and UV wavelengths from Chandra \citep{2000SPIE.4012....2W}, XMM-Newton \citep{2001A&A...365L...1J}, and the Hubble Space Telescope have characterized coronal and chromospheric emissions across a wide range of stellar types \citep[e.g.,][]{2009A&ARv..17..309G,2019LNP...955.....L}. These studies reveal that stellar coronae exhibit temperatures and emission measures comparable to or exceeding those of the solar corona, particularly in rapidly rotating stars and young stellar objects. The observed correlation between X-ray luminosity and rotation rate underscores the role of magnetic dynamo processes in driving high-energy emissions \citep[e.g.,][]{2011ApJ...743...48W}. Despite these advances, stellar observations remain limited by spatial resolution. Active regions, flares, and coronal loops cannot be directly imaged on other stars, which makes constraining their geometry and energy budgets challenging. Moreover, observing stellar EUV flux, which is crucial for unraveling the evolutionary processes of exoplanetary atmospheres, is challenging because of the strong absorption by the interstellar medium. These limitations motivate the use of scaling relations and extrapolations derived from solar data in interpreting stellar activity signatures.

Solar data serve as a cornerstone for modeling stellar magnetic activity. Through the analysis of solar and stellar data, empirical power-law relations have been derived between magnetic flux and X-ray luminosity \citep[e.g.,][]{1998ApJ...508..885F,2003ApJ...598.1387P}. These relations provide a foundation for predicting stellar coronal properties based on magnetic field measurements. Furthermore, observations suggest that faster-rotating stars have stronger magnetic fields and higher luminosity, a trend summarized as Skumanich's law \citep{1972ApJ...171..565S,1981ApJ...248..279P}. This seminal relation states that stellar rotation rate decreases with age as $\Omega\propto t^{-1/2}$, implying that magnetic activity and high-energy emissions decline over time. Subsequent studies have refined this framework by incorporating dependencies on the Rossby number and saturation effects at high rotation rates \citep[e.g.,][]{1984ApJ...279..763N,2011ApJ...743...48W}. Recent studies further show that the rotation-activity evolution includes features such as the intermediate rotation gap and signs of weakened magnetic braking, demonstrating a more complex behavior than the classical Skumanich's law \citep[e.g.,][]{2016Natur.529..181V,2021NatAs...5..707H,2022ApJ...933L..17M,2023ApJ...951L..49C,2025A&A...697A.177S,2025ApJ...982..114M}. These refinements provide a more comprehensive picture of stellar activity evolution and highlight the importance of rotation in shaping magnetic heating processes.

Observations of solar and stellar coronae reveal emission lines from highly ionized species, indicating temperatures of 10$^{6}$--10$^{7}$ K \citep[e.g.,][]{2018LRSP...15....5D}. Two representative scenarios---wave dissipation and nanoflare heating---have been extensively examined in the SDO era \citep[e.g.,][]{2025arXiv250702111A}. Wave-based models propose that Alfv\'{e}n waves excited in the photosphere propagate along the field lines into the corona and dissipate their energy through resonant absorption or turbulence \citep[e.g.,][]{2011ApJ...736....3V,2005ApJS..156..265C}. The nanoflare hypothesis, however, attributes coronal heating to numerous fine-scale reconnection events that intermittently release energy \citep{1972ApJ...174..499P,1988ApJ...330..474P}. Both mechanisms may operate simultaneously, and their relative contributions remain an active area of research.

This paper reviews the recent progress in understanding solar-stellar connections, focusing particularly on the contributions of SDO. We address two fundamental questions: How can we characterize stellar active regions? and do Sun-like stars share a common heating mechanism? The first question concerns the tools that enable us to infer the magnetic properties of active regions on stars despite the limitations of unresolved data. The second question asks whether processes responsible for heating the solar corona, such as wave dissipation and nanoflares, are universal among Sun-like stars or vary with stellar parameters. Through a review of {\it Sun-as-a-star} studies utilizing the SDO data, in which spatially resolved images are integrated to mimic the unresolved stellar data, we aim to show the importance of solar observations in addressing these questions and outline future directions for solar-stellar studies.

The remainder of this paper is organized as follows. Section \ref{sec:ar} reviews Sun-as-a-star studies that characterize stellar active regions using multi-wavelength solar observations and discusses diagnostic tools for interpreting unresolved stellar signals. Section \ref{sec:heating} examines universal atmospheric heating mechanisms, focusing on empirical flux-flux scaling laws, spectral reconstructions, and their applicability to Sun-like stars. Section \ref{sec:discussion} highlights the role of SDO in bridging solar and stellar physics, explores recent applications beyond conventional studies, and discusses future directions. Finally, Section \ref{sec:conclusions} summarizes this paper.

\section{Characterization of Stellar Active Regions}\label{sec:ar}

Just as the area of starspots\footnote{We use ``spot area'' as a convenient shorthand here. However, estimates from unresolved photometry are degenerate with stellar inclination and spot-photosphere contrast. Therefore, terms such as ``spot coverage'' or ``spot filling factor'' may be more appropriate.} can be estimated from the rotational modulation in a star's visible light curve, observing stellar light curves at multiple wavelengths may allow us to infer the stellar active regions and their surrounding magnetic environments. This is the fundamental idea underlying the work of \citet{2020ApJ...902...36T}.

This study is primarily aimed at understanding how solar active regions, when observed as if the Sun were an unresolved star, influence spectral irradiance across multiple wavelengths. This approach is motivated by the need to interpret stellar activity signatures in disk-integrated light curves. The authors examined how different active region types, e.g., isolated sunspots, spotless plages, and emerging flux regions, affect irradiance in visible, UV, EUV, and X-ray bands during their transit across the solar disk to establish diagnostic tools and characterize stellar active regions.

The study employs full-disk synoptic observations from multiple instruments:
\begin{itemize}
\item SDO/HMI: Provides visible continuum and line-of-sight magnetograms for measuring total unsigned magnetic flux.
\item SDO/AIA: Offers EUV imaging across channels sensitive to chromospheric to coronal temperatures.
\item SORCE/Total Irradiance Monitor (TIM) \citep{2005SoPh..230....7R,2005SoPh..230..129K}: Delivers total solar irradiance (TSI).
\item Hinode/X-Ray Telescope (XRT) \citep{2007SoPh..243....3K,2007SoPh..243...63G}: Provides coronal soft X-ray imaging at 10$^{6.9}$ K.
\item GOES/X-Ray Sensor (XRS): Supplies soft X-ray flux for high-temperature diagnostics.
\end{itemize}
Three representative transit objects were selected along with a quiet-Sun condition for reference, as summarized in Table \ref{tab:ar}. All of these events appeared during the deep solar minimum between Cycle 24 and 25.

\begin{table}
\caption{List of transit events analyzed in \citet{2020ApJ...902...36T}.}
\label{tab:ar}
\begin{tabular}{cccc}
\hline
No. & Event & Object & Central meridian\\
\hline
1 & Quiet Sun (reference) & N/A & December 7, 2019\\
2 & Sunspot & NOAA 12699 & February 11, 2018\\
3 & Spotless plage & Return of NOAA 12713 & July 14, 2018\\
4 & Emerging flux & NOAA 12733 & January 24, 2019\\
\hline
\end{tabular}
\end{table}

\begin{figure}
\centerline{\includegraphics[width=\textwidth]{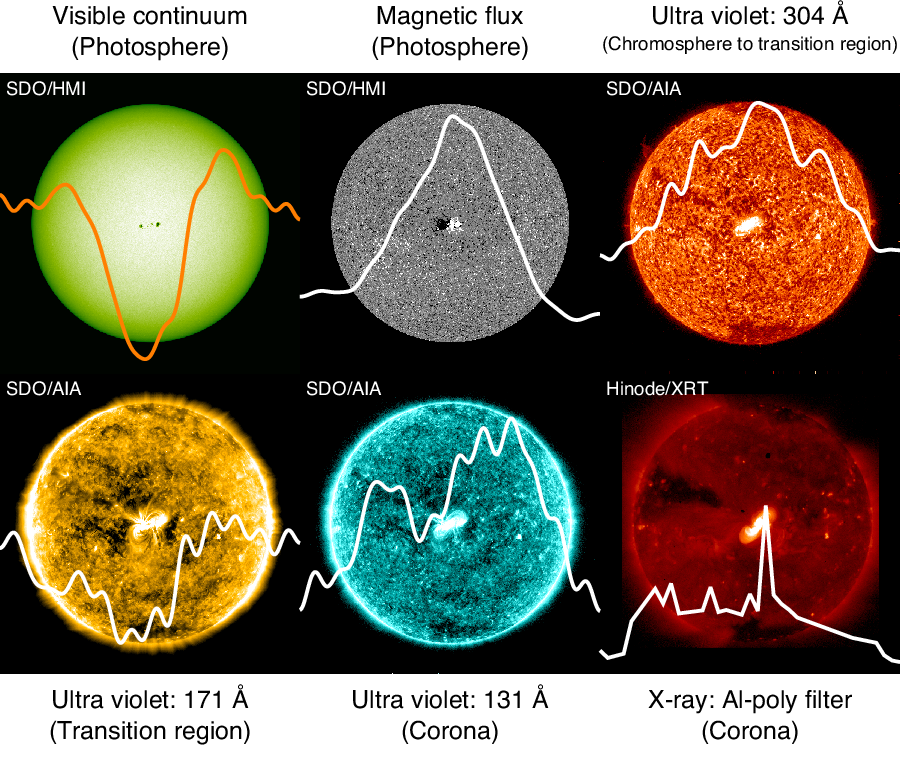}}
\small
\caption{Representative full-disk images of the transiting sunspot (NOAA 12699) with corresponding light curves overlaid. Panels show HMI continuum and magnetogram, AIA 304, 171, and 131 {\AA} channels, and Hinode/XRT. Images were taken at 00:00 UT on 2018 February 11. Figure was created based on \citet{2020ApJ...902...36T}.}
\label{fig:lc}
\end{figure}

For each event, Sun-as-a-star light curves were constructed by integrating full-disk images in multiple wavelength bands (HMI, AIA, and XRT) as the active region transited the disk. Figure \ref{fig:lc} shows a mosaic of the selected full-disk images for the transiting sunspot (NOAA 12699) with the corresponding Sun-as-a-star light curves. These curves include the HMI visible continuum intensity and photospheric magnetic flux, and multi-wavelength observations ranging from UV to soft X-rays by AIA and XRT, respectively.

The key results and important findings are summarized below.
\begin{itemize}
\item {\bf TSI and visible continuum:} The TSI and visible continuum darken when a sunspot is near the central meridian, with brightening occurring near the limb owing to faculae visibility.
\item {\bf Near UV and magnetic flux correlation:} UV bands sensitive to chromospheric temperatures (e.g., AIA 1600 and 1700 {\AA}) strongly correlate with total unsigned magnetic flux, confirming UV brightness as a proxy for magnetic activity.
\item {\bf EUV and X-ray emission trends:} EUV and X-ray fluxes increase with coronal temperature. Light curves exhibit flat-topped profiles during the disk transit, reflecting optically thin coronal plasma.
\item {\bf Spotless plage and emerging flux:} Unlike the sunspot case, spotless plages do not produce visible darkening, whereas emerging flux regions create asymmetries in all light curves, reflecting the temporal evolution of the magnetic field.
\item {\bf Time lags between photospheric and coronal signals:} Because the coronal loops are visible above the limb, EUV and X-ray curves begin before the plage brightening occurs in the visible continuum. Likewise, the EUV and X-ray remain bright even after active regions have rotated off the disk. These time lags suggest loop heights above the active regions and offer a diagnostic for magnetic topology in unresolved stars.
\item {\bf Antiphased EUV variations:} The AIA 171 {\AA} channel, which is sensitive to sub-MK plasmas, sometimes exhibits anti-phased behavior relative to hotter emission lines. That is, the 171 {\AA} light curve shows ``dimming'' when the active region is on the disk.
\end{itemize}
The AIA 171 {\AA} image in Figure \ref{fig:lc} clearly shows a dark area spreading widely around the active region at the disk center, which is distinctly different from the coronal holes seen in the XRT image. Analysis of the differential emission measures across the disk revealed that during the 171 {\AA} dimming, sub-MK plasma decreased significantly around active regions compared to the reference quiet-Sun level, whereas the emission measure of the high-temperature component increased. Similar 171 {\AA} dimming was observed around the newly-emerging and evolved active regions \citep[e.g.,][]{2012ApJ...760L..29Z,2021ApJ...909...57S,2021ApJ...912....1P,2023A&A...680A..61L}.

\begin{figure}
\centerline{\includegraphics[width=0.6\textwidth]{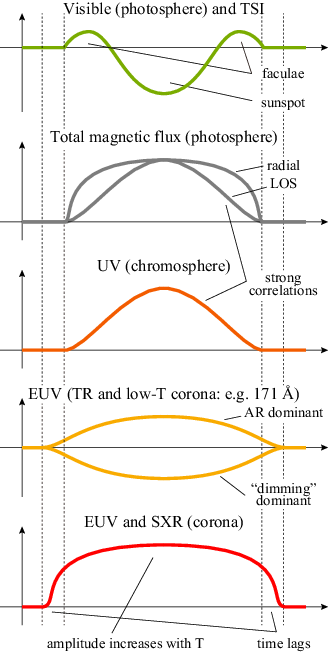}}
\small
\caption{Schematic illustration of light curve variations during active region transits across the disk at different wavelengths. In each panel, the thick solid line shows the temporal variation of irradiance, whereas the reference quiescent level is indicated by the thin horizontal axis. The middle of the curves represents the time when the target active region is at the central meridian. Sunspots produce a central dip in visible and TSI curves; spotless plages remove this dip; and emerging fluxes introduce asymmetry about the central meridian in all wavelengths. Figure is reproduced from \citet{2020ApJ...902...36T}.}
\label{fig:schematic}
\end{figure}

These findings are schematically summarized in Figure \ref{fig:schematic}. Visible and TSI light curves reflect spot and facular effects. The photospheric magnetic flux and chromospheric emissions exhibit high correlations in general, but the total magnetic flux derived from the line-of-sight component shows a stronger correlation with the chromospheric emissions than that derived from the radial component. EUV lines sensitive to the transition region and low-temperature corona showing the ``dimming,'' such as AIA 171 {\AA}, exhibit a dipped profile. Coronal bands show prolonged brightening due to extended loops above the limb, and the time lags between photospheric and coronal signals offer a diagnostic for the loop heights.

Therefore, it was concluded that the long-term monitoring of stars across multiple wavelengths, rather than just the visible continuum, can provide a framework for probing starspots on the surface and characterizing magnetic structures in the upper atmosphere.

The usefulness of multi-wavelength observations for characterizing active regions also has important implications for stellar rotation measurements. Specifically, bluer wavelength bands are more sensitive to facular and plage components, which often persist longer than spots. Thus, this enhanced sensitivity may improve the accuracy of rotation determinations \citep[e.g.,][]{2024ApJ...963..102L}.
\vspace{2cm}

\section{Universality of Atmospheric Heating}\label{sec:heating}

\subsection{Universal Heating in Solar and Stellar Atmospheres}

The challenge of explaining why solar and stellar outer atmospheres, particularly the chromosphere and corona, reach temperatures far exceeding those of the photosphere was introduced in Section \ref{sec:introduction}. Here, we focus on empirical evidence for universal heating laws that govern both the Sun and Sun-like stars. Recent studies using decade-long SDO observations have advanced this field by establishing power-law scaling relations between magnetic activity proxies and spectral irradiance. These relations provide a quantitative framework for interpreting stellar activity and modeling atmospheric heating. Here we examine two studies, that of \citet{2022ApJ...927..179T} and its follow-up study by \citet{2022ApJS..262...46T}, developing universal scaling laws for solar and stellar atmospheres. We also describe the approach of \citet{2023ApJ...945..147N}, which extended these scaling laws to reconstruct high-energy spectra of the Sun-like stars.

\subsection{Scaling Laws Linking Magnetic Flux and Irradiance}

The paper by \citet{2022ApJ...927..179T} aims to determine whether magnetic heating processes follow universal scaling laws across solar and stellar atmospheres. Specifically, it seeks to quantify the relationship between magnetic flux $\Phi$ and irradiance $F$ in different temperature regimes, to assess whether these relations can predict stellar activity levels. This is to expand the previous studies on X-ray flux \citep[$F_{\rm X}$ vs. $\Phi$: e.g.,][]{1998ApJ...508..885F,2003ApJ...598.1387P} to cooler temperatures, i.e., the transition region and chromosphere.

Photospheric magnetic flux was derived from daily magnetograms produced by HMI. For each day, the total unsigned magnetic flux over the visible solar disk was calculated, providing a global measure of magnetic activity. In parallel, Sun-as-a-star spectral irradiance measurements were obtained from multiple instruments: SORCE/XPS for EUV and soft X-ray bands indicative of coronal plasma; SORCE/SOLSTICE for UV wavelengths representing chromospheric and transition region emissions; SORCE/SIM and SOLIS/ISS for visible and near-infrared lines; and GOES/XRS for soft X-ray flux. This multi-instrument approach ensures coverage of a broad range of atmospheric layers from the chromosphere to corona.

The analysis proceeds by correlating daily magnetic flux with corresponding irradiances for various spectral bands. These correlations are examined in double logarithmic space, and linear regression is applied to derive power-law relationships of the form
\begin{eqnarray}
F\propto\Phi^{\alpha},
\end{eqnarray}
or
\begin{eqnarray}
\log{F}=\alpha\log{\Phi}+\beta,
\end{eqnarray}
where the power-law index $\alpha$ represents the efficiency of the atmospheric heating with regard to the surface magnetic flux, and $\beta$ is an offset value. By repeating this procedure across multiple wavelengths, a systematic picture of how $\alpha$ varies with temperature was constructed.

\begin{figure}
\centerline{\includegraphics[width=1.3\textwidth,angle=90]{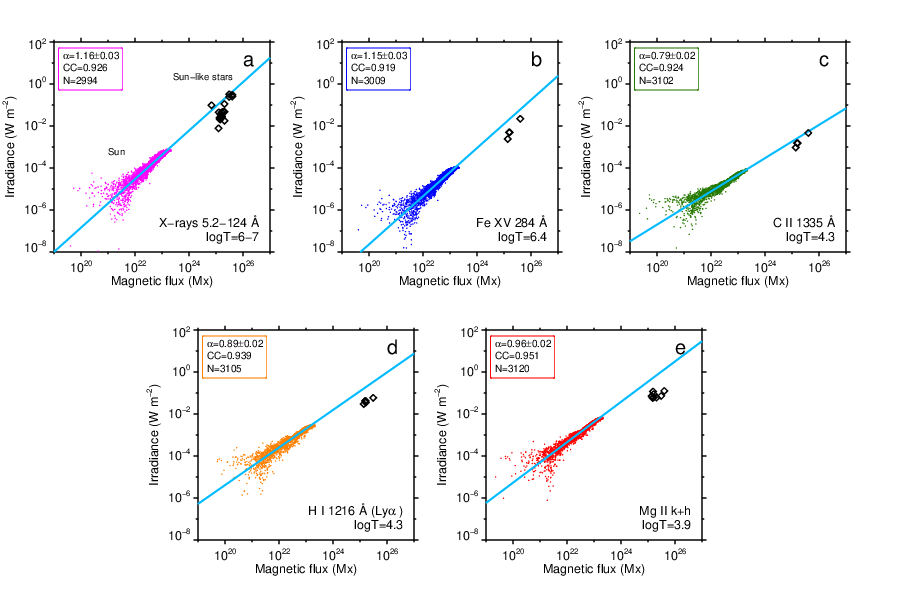}}
\small
\caption{Double-logarithmic scatter plots of solar irradiance versus total unsigned magnetic flux for five selected wavelengths, with power-law fits ($F \propto \Phi^{\alpha}$). Each panel shows the power-law index (slope) $\alpha$, correlation coefficient, and number of data points. Diamonds indicate the stellar data from the literature. Figure is reproduced from \citet{2022ApJ...927..179T}; however, the stellar magnetic flux for the stars was recalculated by correcting the error in the filling factor.}
\label{fig:cc}
\end{figure}

\begin{table}
\caption{Temperatures and power-law indices $\alpha$ against total magnetic flux for various spectral lines from \citet{2022ApJ...927..179T}. The temperatures of optically thick chromospheric lines are in parentheses.}
\label{tab:cc}
\begin{tabular}{ccc}
\hline                     
Feature & $\log{(T/{\rm K})}$ & Power-law Index $\alpha$\\
\hline
X-rays 1--8 {\AA} & 6--7 & $1.42\pm 0.04$\\
X-rays 5.2--124 {\AA} & 6--7 & $1.16\pm 0.03$\\
Fe XV 284 {\AA} & 6.4 & $1.15\pm 0.03$\\
Fe XIV 211 {\AA} & 6.3 & $1.15\pm 0.03$\\
Fe XII 193$+$195 {\AA} & 6.2 & $1.14\pm 0.03$\\
F10.7cm radio & $\sim$6 & $1.24\pm 0.03$\\
He II 256 {\AA} & 4.9 & $1.14\pm 0.03$\\
Si IV 1393 {\AA} & 4.9 & $0.90\pm 0.02$\\
Si IV 1402 {\AA} & 4.9 & $0.83\pm 0.02$\\
C II 1335 {\AA} & 4.3 & $0.79\pm 0.02$\\
H I 1216 {\AA} (Ly$\alpha$) & 4.3 & $0.89\pm 0.02$\\
Mg II k 2796 {\AA} & (3.9) & $0.95\pm 0.02$\\
Mg II h 2803 {\AA} & (3.9) & $0.97\pm 0.03$\\
Mg II k$+$h & (3.9) & $0.96\pm 0.02$\\
Ca II K 3934 {\AA} & (3.8) & $0.87\pm 0.03$\\
Ca II H 3968 {\AA} & (3.8) & $0.86\pm 0.04$\\
H I 6563 {\AA} (H$\alpha$) & (3.8) & $-1.46\pm 0.14$\\
\hline
\end{tabular}
\end{table}

The results are summarized in Figure \ref{fig:cc} and Table \ref{tab:cc}, which reveals clear and physically meaningful trends. Across all wavelengths, irradiance $F$ and magnetic flux $\Phi$ exhibit strong positive correlations, confirming that the magnetic field drives atmospheric heating. However, the slope of the power-law relation, $\alpha$, is not constant but varies systematically with temperature. For coronal emissions observed in EUV and X-ray bands, $\alpha$ is above unity, typically between 1.1 and 1.4, indicating that irradiance increases steeply with magnetic flux in the hottest regions of the atmosphere. This superlinear behavior is consistent with the previous results \citep[e.g.,][]{1998ApJ...508..885F,2003ApJ...598.1387P}. In contrast, chromospheric emissions in the near UV range display much smaller indices, generally between 0.8 and 0.9, suggesting a milder response to magnetic flux enhancement. This contrast in $\alpha$ reflects fundamental differences in heating efficiency: higher-temperature plasma appears to be more sensitive to magnetic energy input, whereas cooler layers exhibit a weaker dependence.

To extend these solar-based scaling laws to the stellar regime, the derived relationships were compared with observations of G-type main-sequence stars ranging in age from 50 Myr to 4.5 Gyr, as summarized in Table \ref{tab:star}. The stellar magnetic field measurements are based on \citet{2020A&A...635A.142K}. Consequently, the stellar data align remarkably well with the solar scaling laws across a wide temperature range, from the corona to the chromosphere, despite significant differences in the power-law indices. This striking consistency provides strong observational evidence that the atmospheric heating mechanism is universal among the Sun and Sun-like stars, regardless of their age or activity level.

\begin{table}
\caption{Characteristics of the stars analyzed in \citet{2022ApJ...927..179T}.}
\label{tab:star}
\begin{tabular}{cccccccc}
\hline                     
HD & Name & Spectral type & $T_{\rm eff}$ & $\log{g}$ & Age & $P_{\rm rot}$ & $R$\\
 & & & (K) & & (Myr) & (days) & ($R_{\odot}$)\\
\hline
1835 & BE Cet & G3V & 5837 & 4.47 & 600 & 7.78 & 1.00\\
20630 & $\kappa^{1}$ Cet & G5V & 5742 & 4.49 & 600 & 9.3 & 0.95\\
39587 & $\chi^{1}$ Ori & G0V & 5882 & 4.34 & 500 & 4.83 & 1.05\\
56124 &  & G0V & 5848 & 4.46 & 4500 & 18 & 1.01\\
72905 & $\pi^{1}$ UMa & G1.5V & 5873 & 4.44 & 500 & 4.9 & 0.95\\
73350 & V401 Hya & G5V & 5802 & 4.48 & 510 & 12.3 & 0.98\\
76151 &  & G3V & 5790 & 4.55 & 3600 & 20.5 & 1.00\\
82558 & LQ Hya & K1V & 5000 & 4.00 & 50 & 1.601 & 0.71\\
129333 & EK Dra & G1.5V & 5845 & 4.47 & 120 & 2.606 & 0.97\\
131156 & $\xi$ Boo A & G7V & 5570 & 4.65 & 200 & 6.4 & 0.83\\
166435 &  & G1IV & 5843 & 4.44 & 3800 & 3.43 & 0.99\\
175726 &  & G0V & 5998 & 4.41 & 500 & 3.92 & 1.06\\
190771 &  & G2V & 5834 & 4.44 & 2700 & 8.8 & 1.01\\
206860 & HN Peg & G0V & 5974 & 4.47 & 260 & 4.55 & 1.04\\
\hline
 & Sun (mean) & G2V & 5777 & 4.44 & 4600 & 25.4 & 1.00\\
\hline
\end{tabular}
\end{table}

Another important finding relates to the dependence on the solar cycle. The authors observed that the power-law indices $\alpha$ vary with activity level. During a solar maximum, $\alpha$ reaches its minimum in most temperature regimes, suggesting that heating efficiency is reduced when surface magnetic flux reaches high levels. This may be due to physical limits in energy transport or changes in magnetic topology that reduce heating efficiency. Conversely, during a solar minimum, $\alpha$ is at its maximum, suggesting that heating efficiency increases.

In summary, the key results of \citet{2022ApJ...927..179T} establish a quantitative link between magnetic activity and atmospheric heating. They reveal temperature-dependent variations in heating efficiency and identify its cycle dependence. Crucially, they demonstrate that these relationships apply to Sun-like stars of all ages and activity levels over a broad temperature range, not only in the corona, as was previously found, but also in the transition region and chromosphere.

\subsection{Catalog of Power-law Indices for Solar Activity Proxies}

Building on the work of \citet{2022ApJ...927..179T}, \citet{2022ApJS..262...46T} transformed the concept of universal scaling laws into a comprehensive catalog. Although the earlier study demonstrated that magnetic flux $\Phi$ and irradiance $F$ follow power-law relationships, $\log{F}=\alpha\log{\Phi}+\beta$, this follow-up paper broadened the scope by systematically analyzing multiple solar activity proxies and a wide range of spectral irradiances. The goal was to create a detailed reference dataset of power-law index $\alpha$ and offset $\beta$ for solar reconstructions and stellar modeling.

The main advancement lies in the diversity of solar activity proxies and spectral coverage. Rather than focusing primarily on magnetic flux, several widely used indicators of solar activity were incorporated, as follows:
\begin{itemize}
\item Total magnetic flux, based on HMI full-disk magnetograms.
\item Sunspot number and area, capturing the size and frequency of active regions.
\item F10.7 cm radio flux, often used as a standard proxy for solar EUV output and coronal activity.
\end{itemize}
On the irradiance side, the dataset was expanded to cover emissions from all atmospheric layers:
\begin{itemize}
\item Chromospheric lines such as Mg II and Ca II formed at temperatures around 10$^{4}$ K.
\item Transition region lines such as C IV and Si IV formed around 10$^{5}$ K.
\item Coronal lines including Fe XII and Fe XV, observed in EUV and X-ray bands, corresponding to 10$^{6}$--10$^{7}$ K plasmas.
\end{itemize}

\begin{figure}
\centerline{\includegraphics[width=\textwidth]{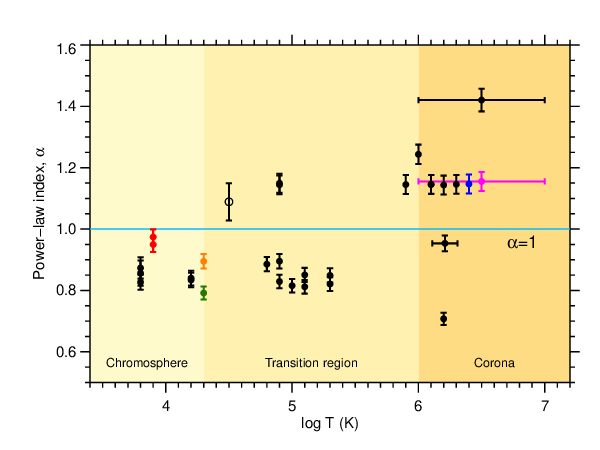}}
\small
\caption{Power-law index $\alpha$ derived from solar data plotted against the formation temperature of various spectral lines. Vertical bars show fitting errors; horizontal bars indicate temperature ranges for three X-ray observations. The $\alpha=1$ level is indicated by a sky-blue line. He I 10830 {\AA}, which shows an anti-phased cycle variation against the solar activity, is indicated by an open circle. Colored symbols denote those compared with the stellar data in Figure \ref{fig:cc}. Figure is reproduced from \citet{2022ApJS..262...46T}.}
\label{fig:pl}
\end{figure}

One of the important discoveries brought about by compiling this catalog is that the power-law index $\alpha$ is sub-unity not only for the chromospheric emissions but also for the transition region lines, as shown in Figure \ref{fig:pl}. This trend suggests that the heating mechanism of the transition region, or at least its efficiency, is similar to that of the chromosphere and less significant compared to that of the corona. Additionally, the catalog enables the reconstruction of solar spectral line fluxes for historical periods using long-term sunspot or F10.7 records. This can improve radiative forcing models in the Earth's and planets' upper atmosphere and on climate variability \citep[e.g.,][]{2010JGRA..11512112K,2017ApJ...836L...3A}. The catalog also enables the prediction of irradiance for F-, G-, and K-type dwarfs based on observable proxies, thereby supporting studies of stellar activity and exoplanetary environments.

\subsection{Reconstructing XUV Spectra of Active Sun-like Stars}

Further extending the work of \citet{2022ApJ...927..179T}, \citet{2023ApJ...945..147N} aimed to develop an empirical method for reconstructing the EUV and X-ray spectra of active Sun-like stars. These high-energy emissions are essential to understanding planetary atmospheric evolution and habitability \citep[e.g.,][]{2019LNP...955.....L}. However, direct EUV observations of stars are nearly impossible because the interstellar medium absorbs EUV photons. This leaves a significant gap in stellar radiation modeling.

\begin{figure}
\centerline{\includegraphics[width=0.7\textwidth]{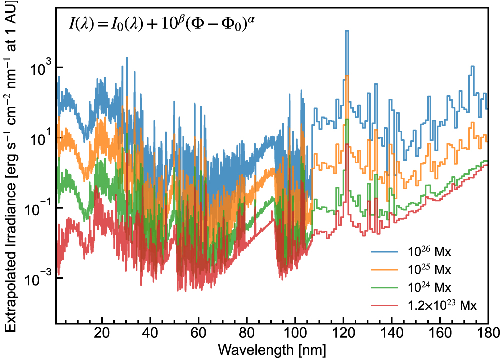}}
\small
\caption{Sample synthesized spectra for stars having a total unsigned magnetic flux of 10$^{24}$, 10$^{25}$, and 10$^{26}$ Mx, derived using Equation (\ref{eq:lambda}). The model spectrum at a solar minimum value of $1.2\times 10^{23}$ Mx is also plotted in red as a reference. Figure is reproduced from \citet{2023ApJ...945..147N}.}
\label{fig:sample}
\end{figure}

To address this issue, the authors propose a method based on solar scaling relations. The key idea is that X-ray and UV (collectively XUV) fluxes strongly correlate with total unsigned magnetic flux. \citet{2022ApJ...927..179T} and \citet{2022ApJS..262...46T} derived power-law relations between magnetic flux $\Phi$ and each line irradiance $F_{\rm line}$ to reconstruct stellar irradiances, whereas \citet{2023ApJ...945..147N} derived power-law relations between magnetic flux $\Phi$ and spectral irradiance $I({\lambda})$ for each wavelength bin $\lambda$,
\begin{eqnarray}
  I(\lambda)=10^{\beta(\lambda)}\Phi^{\alpha(\lambda)},
  \label{eq:lambda}
\end{eqnarray}
to reconstruct entire stellar spectra in the XUV range. Using decade-long data from SDO/EVE, HMI, and other instruments introduced in previous sections, the scaling laws, $\alpha(\lambda)$ and $\beta(\lambda)$, were established for wavelengths ranging from 1 to 1800 {\AA}. The sample spectra for the stars with magnetic flux, $\Phi$, of 10$^{23}$ to 10$^{26}$ Mx are shown in Figure \ref{fig:sample}.

These relations were then applied to reconstruct spectra for three young, magnetically active G-type stars, EK Dra (G1.5V, $\sim$0.1 Gyr), $\pi^{1}$ UMa (G1.5V, $\sim$0.5 Gyr), and $\kappa^{1}$ Ceti (G5V, $\sim$0.6 Gyr), based on the measured magnetic fluxes of these stars. The synthesized XUV and FUV spectra for these stars are publicly available at \url{https://github.com/KosukeNamekata/StellarXUV.git}.

\begin{figure}
\centerline{\includegraphics[width=\textwidth]{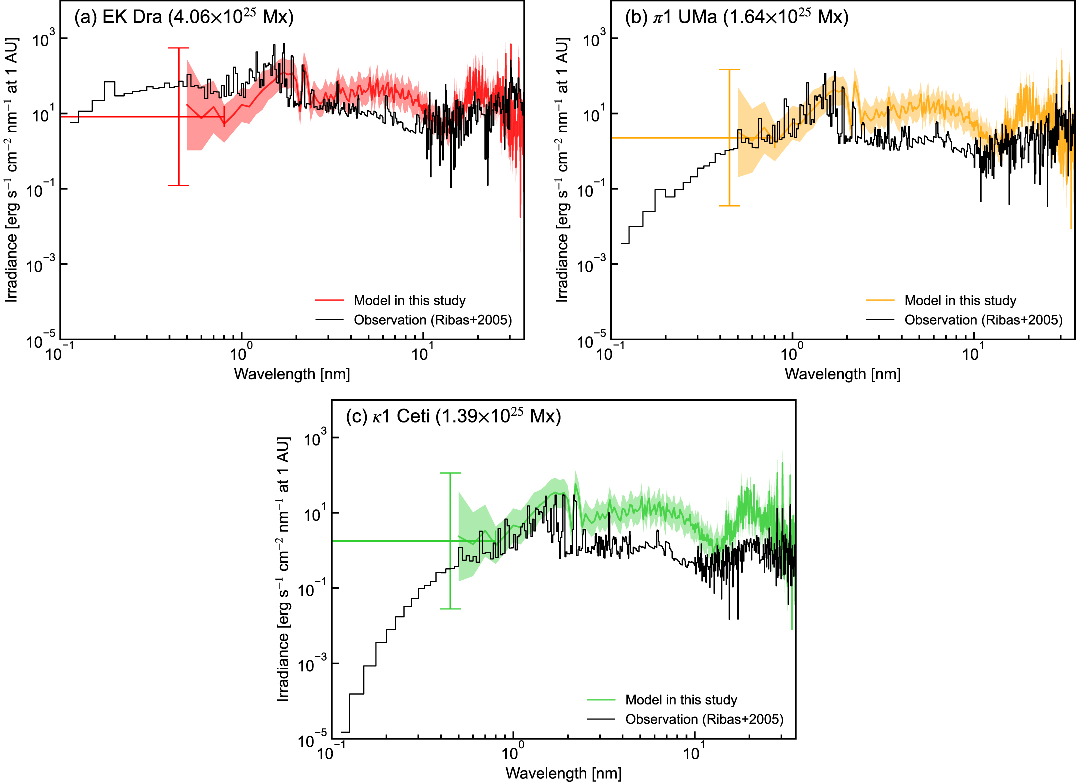}}
\small
\caption{Comparison of observed XUV spectra of active Sun-like stars \citep[black lines, based on][]{2005ApJ...622..680R} with spectra reconstructed from the solar empirical scaling laws (colored lines), with shades indicating uncertainties. Stellar names and their total unsigned magnetic fluxes are noted in each panel. Figure is reproduced from \citet{2023ApJ...945..147N}.}
\label{fig:xuv}
\end{figure}

The reconstructed stellar spectra were compared to actual observations from the ASCA, ROSAT, EUVE, IUE, and Hubble Space Telescope. Figure \ref{fig:xuv} displays the comparisons in the XUV range. The reconstructions are consistent with the ground-truth observations within an order of magnitude. This method successfully reproduces stellar spectra and predicts significantly higher XUV fluxes for young stars than for the present-day Sun. This suggests that planets orbiting these stars undergo intense atmospheric erosion and photochemical changes, which are essential for studying habitability. This approach provides a practical solution for estimating stellar XUV radiation when direct EUV observations are not possible. Thus, it is a valuable tool for exoplanetary research.

\section{Discussion}\label{sec:discussion}

\subsection{The Role of SDO in Bridging Solar and Stellar Physics}

Owing to its unique combination of long-term stability, high-resolution imaging, magnetograph, and broad spectral coverage, SDO has played a pivotal role in enabling Sun-as-a-star studies and solar-stellar associations. These capabilities enabled us to expand beyond solar observations and develop frameworks applicable to Sun-like stars.

Through uninterrupted monitoring by SDO for more than a decade, the full range of solar activity, including the deep solar minimum between Cycle 24 and 25, was captured. This period was particularly important for \citet{2020ApJ...902...36T}, who analyzed active region transits during times of extremely low activity. This extended coverage provides the statistical foundation for flux-flux scaling laws and catalogs in \citet{2022ApJ...927..179T} and \citet{2022ApJS..262...46T}, as well as the baseline for reconstructing stellar XUV spectra in \citet{2023ApJ...945..147N}. Without continuous observations spanning multiple activity phases, identifying the cycle-dependent trends of the atmospheric heating efficiency would have been impossible.

The HMI delivered an unprecedented number of full-disk magnetograms, which enabled precise measurements of magnetic flux over a wide range of solar conditions. The combination of AIA and EVE provided simultaneous observations across EUV and UV wavelengths, covering plasma temperatures from the chromosphere to corona. Combined, these data enabled us to link photospheric magnetism to atmospheric heating signatures. This capability is essential for deriving universal scaling laws that can be extended to stellar regimes and estimating unresolved coronal structures of the stars.

Considering these advantages, SDO data have recently been used in innovative ways that extend far beyond conventional solar studies. AIA and EVE observations have been used to identify coronal dimming signatures as proxies for stellar coronal mass ejections \citep[e.g.,][]{2016SoPh..291.1761H,2021NatAs...5..697V,2025LRSP...22....2V,2025arXiv250719681W}. EVE's Sun-as-a-star spectra have shown the Doppler shifts during flares, offering critical insights into mass motions associated with these events \citep[e.g.,][]{2022ApJ...931...76X,2024ApJ...964...75O,2025ApJ...993..126L}. HMI measurements of white light flares have clarified the continuity of energy-duration scaling between solar and stellar flares, with shorter stellar flare durations than expected from solar extrapolation \citep{2017ApJ...851...91N}.

\begin{figure}
\centerline{\includegraphics[width=0.8\textwidth]{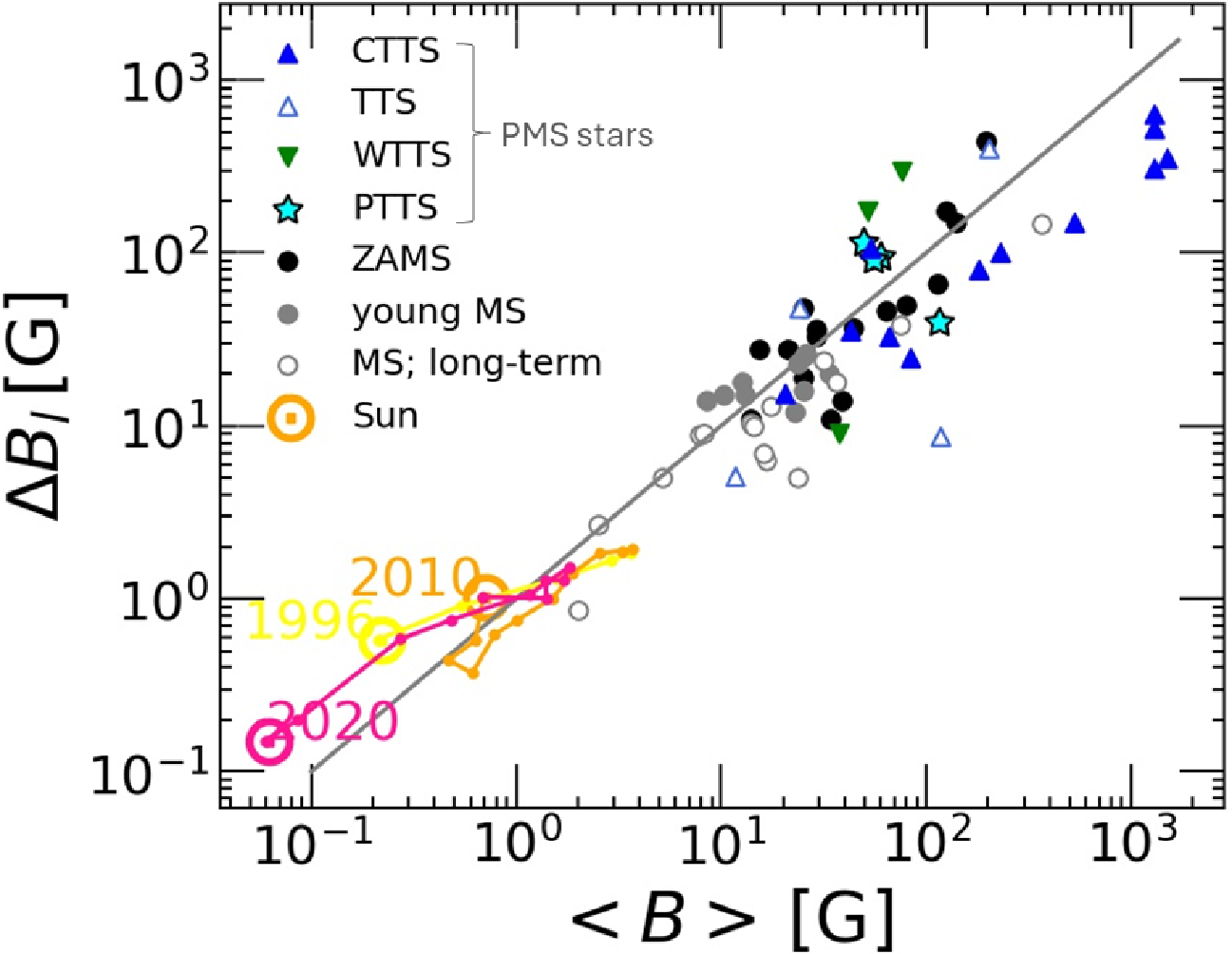}}
\small
\caption{Relationship between the mean magnetic field strength, $\langle B \rangle$, and its variation, $\Delta B$, for the Sun-like stars of different ages. CTTS: classical T Tauri star, TTS: T Tauri star, WTTS: weak-line T Tauri star, PTTS: post T Tauri star, ZAMS: zero-age main-sequence star, and MS: main-sequence star. The solid orthogonal line indicates 1:1. Regardless of age, most stars follow the same scaling law between $\langle B \rangle$ and $\Delta B$. Figure is reproduced from \citet{2025ApJ...985...46Y}.}
\label{fig:pms}
\end{figure}

Beyond flare studies, \citet{2025ApJ...985...46Y} utilized HMI magnetograms and demonstrated a four-order-of-magnitude correlation between disk-averaged magnetic field strength and its variability across pre-main-sequence, zero-age main-sequence, and main-sequence stars, including the Sun (see Figure \ref{fig:pms}). Furthermore, \citet{2024ApJ...965..170S} showed that solar microwave fluxes correlate tightly with the HMI-derived magnetic flux, and that these relations can be extrapolated to at least one nearby Sun-like star ($\epsilon$ Eri), providing an additional solar-stellar diagnostic in the radio bands.

AIA and HMI data have supported exoplanet transit studies by modeling stellar contamination and variability, thereby improving the accuracy of transit depth measurements \citep[e.g.,][]{2015ApJ...802...41L,2016ApJ...817...81L,2020MNRAS.493.5489M}. Sun-based spot-facula models have been used to interpret rotational modulation and reduce systematic errors in stellar activity in transit photometry and radial velocity measurements \citep[e.g.,][]{2016A&A...589A..46S,2016MNRAS.457.3637H,2022ApJ...935....6H}. These studies emphasize the limitations of single-band spot area estimates and advocate for more reliable filling-factor and multi-wavelength methods, where HMI and AIA data provide the basis for decomposing TESS/Kepler light curves \citep[e.g.,][]{2022AJ....163..272E,2025A&A...703A.187L}. Furthermore, HMI-driven statistics have refined the empirical links among rotation, magnetic activity, and dynamo behavior, improving the extrapolation of solar diagnostics to unresolved stellar observations \citep[e.g.,][]{2025A&A...693A.262Z}. Together, these applications demonstrate the versatility of SDO as a bridge between solar and stellar astrophysics.

\subsection{Future Directions}

Although \citet{2020ApJ...902...36T} laid the groundwork for characterizing stellar active regions using multi-wavelength long-term monitoring, several directions can enhance its scope and relevance. First, the analysis is limited to three active region cases, which constrains statistical significance. Expanding the sample to include diverse morphologies and solar cycle phases would improve the robustness of scaling laws and their applicability to stellar contexts. Second, the study focuses on quiescent transits, omitting flare events and other transient phenomena. Incorporating flare-associated irradiance signatures, brightening and coronal dimming, can both offer insights into short-term variability and its impact on stellar light curves. While the possibility of superflare-scale eruptions on Sun-like stars has been discussed in the context of extremely large active regions \citep[e.g.,][]{2013A&A...549A..66A,2013JSWSC...3A..31C,2013PASJ...65...49S,2013ApJ...773..128T,2017ApJ...834...56T}, recent Sun-as-a-star analyses using H$\alpha$ spectral imaging have demonstrated the capability to detect diverse solar active events in spatially integrated data, providing new diagnostics for stellar activity \citep[e.g.,][]{2022NatAs...6..241N,2022ApJ...939...98O}. Third, forward modeling of Sun-as-a-star light curves using numerical simulations for various active region configurations would be valuable in providing theoretical templates for interpreting stellar observations \citep[e.g.,][]{2017ApJ...850...39T,2019ApJ...886L..21T}.

For \citet{2022ApJ...927..179T}, the next priority is expanding stellar diversity. Their study confirmed universal scaling laws using limited main-sequence G-type dwarf stars; however, a broader sample covering different spectral types, ages, and magnetic activity levels (i.e., F-, G-, K-, and M-stars as well as their pre-main-sequence stars) is essential to test the robustness of these laws. Observations from future missions such as Habitable Worlds Observatory, Athena, LAPYUTA, ESCAPE and other facilities will be crucial in this effort. Although the observed superlinearity ($\alpha>1$) of coronal emissions was reproduced by the 1D coronal loop simulation of \citet{2021A&A...656A.111S}, the sublinearity ($\alpha<1$) of chromospheric and transition region emissions has not been examined in depth by numerical modeling \citep[for observations, see, e.g.,][]{1975ApJ...200..747S,1989ApJ...337..964S,2009A&A...497..273L,2018A&A...619A...5B}. Realistic simulations employing radiative magnetohydrodynamics may provide deeper insight into the mechanisms behind such nonlinear behaviors \citep[e.g.,][]{2011A&A...531A.154G,2017ApJ...834...10R,2017ApJ...848...38I,2019NatAs...3..160C}. Recently, \citet{2024NatAs...8..697B} discovered that spatially resolved solar data do not show a similar correlation on small spatial scales. They claim that small-scale magnetic braiding is key to heating the atmosphere. Thus, future work should also focus on scale dependence.

Existing approaches to reconstructing stellar XUV line intensities and spectra are the inversion based on differential emission measures \citep[e.g.,][]{2011A&A...532A...6S,2021ApJ...913...40D,2025AJ....170..159F} and the extrapolation from the Ly-$\alpha$ measurement based on empirical flux-flux scaling laws \citep[e.g.,][]{2014ApJ...780...61L}. However, both of these methods require costly satellite observations to directly measure XUV. The new approaches of \citet{2022ApJS..262...46T} and \citet{2023ApJ...945..147N}, which estimate XUV lines and spectra from observed magnetic fluxes, utilize ground-based telescopes and have the advantage of not requiring satellite observations. Expanding the sample beyond EK Dra, $\pi^{1}$ UMa, and $\kappa^{1}$ Ceti to include stars with diverse rotation rates and magnetic activity levels can test the universality of the method. Notably, numerical modeling plays a crucial role in supporting these reconstructions by providing physically consistent XUV spectra \citep[e.g.,][]{2024A&A...691A.152S}. Incorporating flare contributions and time-dependent variability can further improve accuracy.

Combining multiple reconstruction methodologies can create a unified scheme for predicting stellar radiation environments across a wide range of conditions.

\section{Conclusions}\label{sec:conclusions}

The SDO satellite has provided continuous, high-resolution observations that have transformed our understanding of the solar-stellar connection. By providing uninterrupted monitoring for more than one solar cycle, SDO has enabled studies linking surface magnetic activity to atmospheric heating over a wide temperature range. Combining precise magnetograms from HMI and broad spectral coverage from AIA and EVE enabled us to quantify the correlations between magnetic flux and irradiance from the chromosphere to corona. These data revealed systematic trends in heating efficiency, leading to scaling laws applicable not only to the Sun but also to Sun-like stars.

Integrating spatially resolved solar observations into Sun-as-a-star datasets provides essential benchmarks for decoding unresolved stellar active region signals. The multi-wavelength capability of SDO has led to catalogs of power-law indices and methods for reconstructing stellar XUV spectra, offering practical solutions to challenges such as interstellar absorption. SDO data have recently supported brand-new applications beyond atmospheric heating, including stellar CME synthesis and exoplanet transit modeling.

In summary, SDO's capabilities have bridged solar and stellar physics, enabling predictive models of stellar activity that advance our understanding of how magnetic processes shape stellar atmospheres, impacting planetary environments and their habitability.

\begin{acks}
The author thanks Yuta Notsu and Kosuke Namekata for their valuable comments on the manuscript.
The author is grateful to the anonymous referee for the thoughtful and constructive feedback, which significantly improved the quality of this manuscript.
This work was supported by JSPS KAKENHI grant Nos. JP21H04492 (PI: K. Kusano) and JP25K01041 (PI: K. Namekata).
\end{acks}

\bibliographystyle{spr-mp-sola}
\bibliography{manuscript}

\end{document}